**What might matter in autonomous cars adoption: first person versus third person scenarios**

Authors: **Eva Zackova** (zacka@ntc.zcu.cz), **Jan Romportl** (rompi@ntc.zcu.cz)

Affiliation: Man-Machine Interaction Dpt., New Technologies – Research Centre, University of West Bohemia in Pilsen, Czech Republic

Corresponding author: Jan Romportl, rompi@ntc.zcu.cz, New Technologies – Research Centre, University of West Bohemia, Univerzitni 8, 306 14 Plzen, Czech Republic

**Abstract**
The discussion between the automotive industry, governments, ethicists, policy makers and general public about autonomous cars' moral agency is widening, and therefore we see the need to bring more insight into what meta-factors might actually influence the outcomes of such discussions, surveys and plebiscites. In our study, we focus on the psychological (personality traits), practical (active driving experience), gender and rhetoric/framing factors that might impact and even determine respondents' a priori preferences of autonomous cars' operation. We conducted an online survey (N=430) to collect data that show that the third person scenario is less biased than the first person scenario when presenting ethical dilemma related to autonomous cars. According to our analysis, gender bias should be explored in more extensive future studies as well. We recommend any participatory technology assessment discourse to use the third person scenario and to direct attention to the way any autonomous car related debate is introduced, especially in terms of linguistic and communication aspects and gender.

**Keywords**
autonomous cars; a priory acceptance of technology; trolley problem; Big Five; Temperament and Character Inventory test

**Acknowledgement**
We would like to thank our colleague Daniel Soutner for literature recommendations, and the Institute of Mental Health in the Czech Republic for providing guidelines for personality traits tests implementation.

**Funding**
The research leading to these results has received funding from the Norwegian Financial Mechanism 2009-2014 and the Ministry of Education, Youth and Sports of the Czech Republic under Project Contract no. MSMT-28477/2014, Project no. 7F14236.

**1. Introduction**

The current debate on autonomous cars (AC) is growing immensely across various disciplines ranging from machine engineering, computer science and artificial intelligence through design, ergonomics, public health and responsible research and innovation area to economy, ethics and philosophy. More than dozen manufacturers are currently on the market developing either partially or fully automated vehicles (AV), and it is plausibly expected that more are willing to join this promising area soon (Endsley 2017, KPMG 2016). As the major automobile companies are already convinced that the future of transportation lies in AV, the general public as well as policy makers and scholars are not so united in their attitude, expressing their worries and doubts about the safety and reliability of self-driving cars.

While the technical problems seem to be just a matter of time today, the governance, policy and legal issues persist and naturally dominate the discussions of AV implementation into our everyday life of commuting and transportation, as these are watched with high interest of public that is usually very hard to be brought to a consensus.

As even broader discussion of the AC on the European level can be expected in coming years, the expectations, presumptions and overall feelings toward AC are already being analysed on the local and international markets (Volvo Car USA 2016, Bansal & Kockleman 2017, Kyriakidis et al. 2015, Payre et al. 2014). From the perspective of ethics and responsible research and innovation (RRI), many experts call for action in order to assess pros and cons of this emerging technology in terms of public health and safety, ethical, legal and social issues as well as to strengthen the public engagement in the running discussions (Goodall 2014, Fagnant and Kockleman 2015, Hevelke & Nida-Rümelin 2015, Gogoll & Müller 2016, Fleetwood 2017, Geistfeld 2017). Moreover, in February 2017 European Parliament adopted[1] a study titled *European Civil Law Rules in Robotics* (Nevejans 2017) to begin a process of elaboration of concrete policy and legislation in area of autonomous robots, including AC. A key subject of this study is the possibility to establish robots as legal persons; a concept of electronic persons has been introduced in the study, opening new perspectives regarding the liability for AC's accidents and damages.

In the Czech Republic, the debate on autonomous mobility has been fostered recently by news released by the biggest Czech automobile manufacturer Škoda Auto, claiming in public media that they are preparing to step in the area of AC industry as well. In addition, the government has expressed support and willingness for investments into the development of autonomous mobility infrastructure for testing in the Czech Republic before 2020 (ITS 2015).

As far as the authors are aware of, there has been no study published yet related to attitudes towards AC specifically in the Czech Republic. From this point of view, our paper represents the first attempt to shed more light on basic preferences of self-driving car's behaviour in Central European Country of Czechia.

Since the discussion between the automotive industry, governments, ethicists, policy makers and general public is spreading, we see the need of bringing more insight into what meta-factors might actually influence the outcomes of such discussions, surveys and plebiscites as especially important. Based on prior research we present in this paper and the fact that majority of studies in area of adoption of AC are focusing on technology acceptance models and user preferences in terms of perceived usefulness, comfort, safety, travel cost, time saving etc. (van der Berg & Verhoef 2016, Ro & Ha 2017, Milakis et al. 2017), we decided to focus on the psychological (personality traits), practical (active driving experience) and rhetoric factors (biases coming from the way the question about AC is formulated) that might impact and even determine respondents' a priory preferences of AC's behaviour. We also include gender and education into our analysis as these are very common factors in any decision-making process. As shown by few other studies, especially the psychology of potential AC's users is something worth to investigate in detail (Kyriakidis et al. 2015, Hohenberger et al. 2017).

## 2. Related research

### 2.1 Ethics and policy of autonomous cars

Why the autonomous cars pose so much moral responsibility? Well, the main reason is that their arrival rises the ultimate question of life and death, literally. As we all know, accidents happen when people are driving cars, and even though the autonomous cars promise to decrease the rate of fatal car incidents and lethal injures, basically everybody take for granted that some (maybe very few indeed) will still happen. In case of human driver of regular or semi-autonomous car, the control of what is going to happen during a car incident usually depends on pure chance and luck (or lack of it). Human cognition is not able to process the information about what is going on quickly enough and taking desirable action in a second or two before the crash. In this regard, human beings are equal. However, the computational power of AC exceeds the human cognitive limitation many times, allowing to implement a complex decision function for the cases of unavoidable car crashes. The main question is whose life should be protected in the first place and under which circumstances. This of course opens a great space for misuse, inequalities and biases in design of AC's inner logic as well as it uncovers our own moral attitudes towards the others and the incredibly complicated issues of moral agency of self-driving cars in general.

---

[1] To see the adoption procedure visit
http://www.europarl.europa.eu/oeil/popups/ficheprocedure.do?lang=en&reference=2015/2103(INL)#document Gateway.

Together with Patrick Lin (2016, p. 82), we believe that one of the biggest problem in regard to AC might be the fact of unpredictable risks and lack of imagination. Experts have already thought through immense AC interaction scenarios but as Lin points out, this still resembles more of thinking about electricity as a substitution for candles, or human car driver for a horse than an actual analysis of future autonomous transportation. Most probably, we are not able to imagine future adequately at all. With so many unknown problems waiting for us, Lin advocates for pushing the industry to take the ethics more seriously as this is the only way to compel manufacturers to deliver as technologically mature and reliable AC as possible, lowering the rate of life or death situation and related ethical dilemmas to minimum.

In view of the fact that the fully autonomous driving is not yet available for the end-users and the common experience with this technology is very limited (with few exceptions: Navya Technology 2017a-b, PostBus Switzerland Ltd 2017), the ethical issues of AC are often considered via thought experiments. Namely via the analogy of the so-called trolley problem or trolley dilemma (for details see Otsuka 2008, Thomson 1985, Swann et al. 2010), that has been originally introduced in a very different context by Philippa Foot in 1967. It is believed that the trolley problem thought experiment represents two different ethical stances: the consequentialist (or the utilitarianist) or the non-consequentialist (the deontologist) approach, and are often used rather as a mere test in which category one's opinions fall with no further indication what this might actually mean from a practical point of view. Gerdes & Thornton (2016) provide interesting analysis of these two ethical frameworks and their usability in AC ethics leading to fusion of both as a more practical approach. However, it is not our aim to discuss these ethical frameworks per se.

Various versions of the trolley problem have been developed so far as well as their analogies transformed into the autonomous cars scenarios. These have been discussed broadly in many other studies showing the extensiveness of possible conditions that might play a crucial role in decision-making of AC's users (Rahwan et al. na, Gerdes & Thornton 2016). The most comparable AC scenario is the one, when you have to decide whether your vehicle should strike five people crossing the road or strike the one standing nearby on the pavement. Considering that these two scenarios (trolley dilemma versus autonomous cars dilemma) differs just in the scenery, more or less similar results in terms of utilitarianist versus deontologist ethics would probably occur for the same respondents (although we did not find any study confirming this thesis yet). Based on this presumption, we decided to compare the traditional trolley problem scenarios (track switch and big man) with a modified version of AC problem that is not a 100% pure analogy to them. The main motivation was to see whether there are any correlations between the trolleys and cars scenarios that would predict respondents' answers in the latter. Given that quite a lot of trolley problem surveys have been conducted in experimental philosophy and ethics so far, it might be a game changer to use them retrospectively to predict certain socio-demographic groups' attitudes towards various AC dilemmas.

## 2.2 Adoption of autonomous cars

However much interesting and even disrupting the ethical dilemmas of AC behaviour are, the AC industry and technology assessment field are maybe more concerned about the acceptability and adoption predictors of AC themselves (KPMG 2016, Bansal & Kockelman 2017). For example, Volvo Car USA (2016) presented last year results of their own *Future of Driving Survey* focused on AC adoption in the United States. The main conclusions they make are quite optimistic in regards to how much people trust the AC in terms of safety. In average, 69% of people across the nation believed AC will keep them safer and 68% believed use of AC would lower the number of traffic accidents. At the same time, a majority of respondents (90%) thought that governments were slow in their policy and regulatory respond to AC. This Volvo's press release highlight the fact, that 68% of respondents want to preserve the freedom of decision when to drive manually and when to switch to the autopilot mode. Surprisingly, the reason for that does not lie in the safety aspect, but in perceived joy from driving, that is regarded by respondents as a luxury.

These Volvo's results about the attitudes towards AC are in accordance with the study conducted in France few years earlier that also reported mainly positive attitude of the respondents (all the respondents were drivers) towards AC adoption (Payre et al. 2014). It focused on intention to use AC without prior direct experience, looking for the underlying aspects and possible intention predictors such as a priori acceptability, personality traits and behavioural adaptation. According to their data, approximately two thirds of respondents were in favour of AC use. Furthermore, high score in sensation-seeking personality trait strongly correlated with higher intention to use AC. However, the authors noted this should be interpreted with cautious as sensation-seeking might be in contradiction with pleasure of driving, that might completely diminish the effect of sensation-seeking once the AC are really used and commonly spread. As the Volvo's (2016) study shows, the pleasure of driving is really something that AC's users are likely to keep in the future.

Volvo Car USA (2016) also announced they want to offer a direct experience of travelling in their AC to one hundred people during 2017, which might shed more light on potential consumers' attitudes and preferences as well. However, no further details were given and it is not clear whether this is entirely matter of marketing strategy or there is a plan to perform it as some kind of naturalistic study similar to Endsley's (2017) research based on real (non-simulator) environment experience with Tesla SA 70 autonomous car. Endsley's preliminary study represents a very first attempt to study a typical AC and human interaction under real conditions and even though the results cannot be yet generalized, he suggests great portion of recommendations for both naturalistic and simulation studies with more subjects as well as for future design of the AC's interface and even for policy and public-related issues. Another study based on psychological and social context focuses on mutual impact of anxiety, anticipated benefits and need for self-enhancement and their effect on willingness to use AC (Hohenberger et al. 2017). According to the authors' analysis, anxiety decreases the willingness to use the AC while higher level of perceived benefits increases it, which is quite plausible. However, they also show that high level of anxiety can completely erase the positive effects of benefit perception on the willingness to use AC, especially for people with low and moderate need for self-enhancement.

A standardized personality test Big Five, that has been used in our study as well, has been implemented into an international study of Kyriakidis et al. (2015). Regarding psychological factors, this study provides insight into a relationship between neuroticism and agreeableness and readiness to data transmission within the AC grid. People with higher score in neuroticism are less comfortable with it, whereas people with higher score in agreeableness are slightly more comfortable with it.

In regard to the meta-factors in AC's adoption, we found the study conducted by Costa and his colleagues (2014) aimed at native versus foreign language as especially important and complementary to our own findings. In their paper, they present data showing that there is a significant difference in responses to trolley dilemma when given to the participants in their mother tongue and when the same is presented in a foreign language, even though participants know the foreign language and understand the meaning perfectly. In case of dilemma presented in the foreign language, the decisions were more frequently utilitarian (than in case of presentation in mother tongue) which authors explain by the fact that the foreign language presentation reduces emotional and intuitive reactions. This is one of the very few studies showing the importance of pragmatic level in communication of such issues to the public as well as possible limitations of international surveys on AC allowing participation of non-native speakers' respondents (as in case of *Moral Machine* project referred hereinafter).

When speaking of acceptability and adoption of AC, a study published by Bonnefon et al. (2016) is probably one of the most cited recently. They concluded that even though their respondents support the idea of AC that sacrifice their passengers to save others involved in an accident, respondents would prefer not to travel in such vehicles. Moreover, they showed that respondents would also not approve laws ordering self-sacrifice, and such laws would make them less willing to buy an autonomous car. Such results show the limits of our moral attitudes as well as much more complex nature of similar issues that we should approach very carefully, especially in terms of making any early conclusions based on public surveys or experimental philosophy, as these might not reflect the real situation adequately and in full scope. Boneffon and his colleagues continue their research via online MIT project called *Moral Machine* that enables any English-speaking visitor to take a survey on great number of various AC scenarios and compare his/her results with other respondents (Rahwan et al. na). The *Moral Machine* project also serves as an open public platform for broader discussion of machine ethics in general.

## 3. Materials and methods

We conducted our on-line survey from June 2015 to April 2016 using Qualtrics platform. The respondents were recruited via social media, web advertisements, broadcast, and e-mail sent to students and employees of our university. The whole survey was in Czech language. Besides the trolley dilemma and AC dilemma scenarios, respondents provided their basic socio-demographic information, underwent two standardised personality traits tests; the Five Factor Model (Digman 1990), also known as Big Five, and the Temperament and Character Inventory test (TCI; Cloninger 1994); and they also answered whether they are active drivers or non-drivers. We motivated the respondents to participate in the survey by giving them their psychological tests' scores right after they finished the tests.

### 3.1 Sample

After cleaning our data from those who were not eligible due to age and early drop out from the survey, the total number of respondents was 430 (41% women, 59% men). This includes only those respondents that finished the socio-demo questions, active driving question, trolley dilemma and AC dilemma at least. Their median age was 25 (average 28). The total number of respondents that finished the whole survey (i. e. including the two personality traits tests) was 363 (41,8% women, 58,2% men) with median age 25 (average 28.7). All participants were from the Czech Republic, mostly from the western part of the country where we were able to promote the study more intensively. In comparison to Czech average, our sample deviates from national average of higher educated people, i.e. graduated and with doctoral degree (49% out of 363 respondents with full response, 47% out of 430 with partial response versus 24.5 % as the overall national average in 2015 according to Eurostat).[2]

---

[2] http://ec.europa.eu/eurostat/statistics-explained/index.php/File:Share_of_the_population_by_level_of_educational_attainment,_by_selected_age_groups_and_country,_2015_(%25).png

## 3.2 Problem statement and hypotheses

As other studies showed (Kyriakidis et al. 2015, Hohenberger et al. 2017, Costa et al. 2014), various meta-factors and biases such as psychological traits, language, social status etc. might be crucial for public debates on autonomous cars' ethics, legislation and acceptability. However, we perceive a lack of such studies in current AC technology assessment research where surveys emphasizing perceived benefits such as crash incidents and traffic jams reduction, usefulness, comfort, travel cost efficiency or time saving as adoption predictors prevail.

In our study, we aim to investigate potential biases that might be easily overlooked in such studies and current (and future) public debates, including those using participatory methods, mapping the public attitudes towards coming autonomous cars. Namely, we focus on biases related to gender and level of education, personality traits, active driving experience and biases coming from the way the question about AC is formulated. We also tried to explore the trolley dilemma outcomes as predictors of outcomes in subsequent AC scenarios.

First, all the respondents were given two versions of trolley problem (please see the Appendix for exact formulations). Version A is the classical one with the option to switch the rail track to kill one person instead of five; version B is the big man scenario with the option to push a big man from the bridge to the rail track and get him killed while saving five other lives. In addition, we asked the respondents to evaluate the level of difficulty of the decision they have just made (ranging on 5-point scale from *very easy* to *very difficult*).

In the next step, each respondent was randomly given just one of two versions of autonomous cars problem (please see the Appendix for exact formulations), i.e. there are two disjoint groups of respondents allowing to compare the impact of the way the dilemma is formulated on the outcomes. One group was answering the first-person perspective version of the dilemma (aka „sitting" scenario) being instructed to imagine themselves as a passenger of an AC which is about to be involved in an accident. The other group was answering the third-person perspective version (aka "voting" scenario) being instructed to imagine themselves as voting in a plebiscite about the inner logic that should be implemented in AC. The decision is about the same situation as in the first version, but this time the respondent is not situated in the car, instead he/she is voting about the general consensus whether AC should kill their passengers or rather pedestrians in such situations. In both of these versions, we also asked the respondents to evaluate the level of difficulty of the decision they have just made (ranging on 5-point scale from *very easy* to *very difficult*).

Our hypotheses are following:

**H1** There are associations between the psychological traits (TCI and Big Five values) and the decisions on the autonomous cars dilemmas and the rating of their respective decision difficulties.

**H2** Strongest association will be with Harm Avoidance (HA), Cooperativeness (CO) and Self-Transcendence (ST) in TCI, and Agreeableness (P, from Czech "Privetivost") in Big Five.

**H3** Respondents who are active car drivers will decide differently than non-drivers. We expect that active drivers tend to prefer their own safety to the prejudice of strangers outside the vehicle.

**H4** Women will decide differently than men. We expect that women will be less likely to kill a stranger/pedestrian.

**H5** More educated people will more frequently prefer killing themselves in the "sitting" version and will have more difficult decision making in the "voting" version.

**H6** The "sitting" scenario will be more difficult to decide than the "voting" scenario.

**H7** There will be a strong association between the trolley problem answers and the autonomous cars answers.

**H8** It will be possible to statistically predict the autonomous cars answers from the other variables (socio-demo and psychological).

## 4. Results and Discussion

A detailed step-by-step analysis (including all the R scripts) of collected data with the responses of participants of the survey are available in a supplementary document of this paper. Here we summarize the results in relation to the aforementioned potential biases expressed by the hypotheses and other important findings discovered during the analysis.

### 4.1 Gender

There is a moderately significant association (p = 0.057 in Fisher test) between gender and the 1-st person "sitting" dilemma outcome (see Table 1): women would prefer to kill themselves more often than men (with odds ratio 1.94). However, there is no such a difference in the 3-rd person "voting" scenario where women and men vote in a similar way (see Table 2). This partially confirms our hypothesis (H4) showing that the gender bias manifest itself in the more suggestive 1-st person perspective scenario.

| Gender / Kill whom | Me | Stranger | Total |
|---|---|---|---|
| Female | 24 | 45 | 69 |
| Male | 23 | 84 | 107 |
| Total | 47 | 129 | 176 |

*Table 1: Contingency table of the respondents' gender with respect to the dilemma outcome in the 1st person "sitting" scenario.*

| Gender / Kill whom | Driver | Pedestrian | Total |
|---|---|---|---|
| Female | 15 | 68 | 83 |
| Male | 23 | 81 | 104 |
| Total | 38 | 149 | 187 |

*Table 2: Contingency table of the respondents' gender with respect to the dilemma outcome in the 3rd person "voting" scenario.*

There is a significant (p = 0.0351) difference in how difficult it is for men to decide in the 1-st person "sitting" scenario: being a man decreases the odds of giving higher difficulty by 0.56 than being a woman. It is easier for men to decide whether to kill themselves or a pedestrian. The fact that men tend to let a pedestrian being killed more often than women does not seem to play any difference here. Significance of lower difficulty for men remains even when controlling for the variable of the decision outcome.

There is no significant association between gender and the 3-rd person "voting" scenario difficulty rating. This result is very different from the 1-st person "sitting" scenario where gender makes a difference in the difficulty rating. We can conclude that it is generally easier for men (in comparison with women) to make a decision when they are deciding in the 1-st person scenario which involves the potentiality of killing themselves. When deciding in the 1-st person "sitting" scenario, there may be a gender-based bias (it is more difficult for women to decide). When deciding in the 3-rd person "voting" scenario, there may be a driver-based bias (it is more difficult for drivers to decide).

These results (see Figure 1) are related to hypothesis (H6) that is actually not confirmed here as we cannot say that the 1-st person perspective poses a higher stress (the difficulty level) on the decision than the 3-rd person perspective in general. However it is clear that for men to decide the 1-st person "sitting" scenario is less difficult than for women.

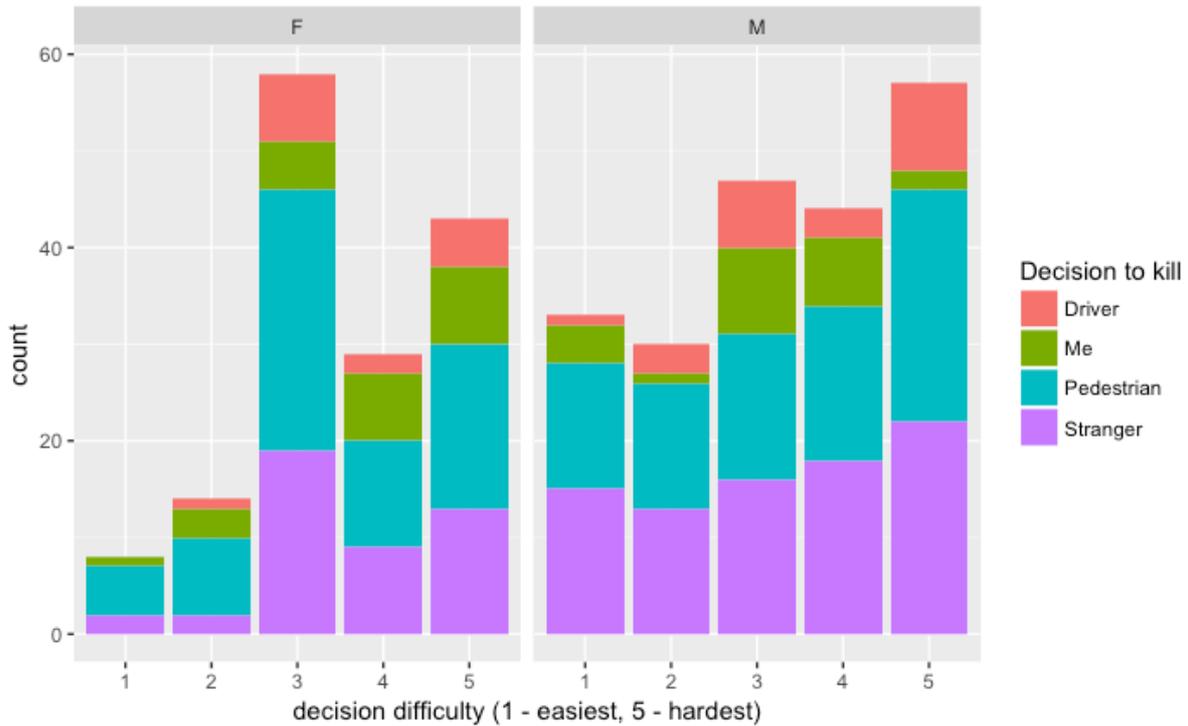

*Figure 1: Bar chart of the decision outcome counts with respect to the decision difficulty ratings and gender in both scenarios.*

**4.2 Active driver**

There is no significant difference between drivers and non-drivers in the AC dilemma outcome, both in the 1-st person "sitting" as well as 3-rd person "voting" scenarios (see Table 3 and Table 4) which is in contradiction with our expectation expressed by hypothesis (H3). At the same time, there is no significant association between being a driver vs. non-driver and the 1-st person "sitting" scenario difficulty rating which again partially contradicts hypotheses (H6).

There is a moderately significant ($p = 0.064$) difference in how difficult it is for active drivers to decide in the "voting" scenario as opposed to non-drivers: being an active driver increases the odds of giving higher difficulty by 1.712 than being a non-driver. This means that it is more difficult for drivers to decide in the 3-rd person "voting" scenario than for non-drivers. Although this was not in our initial list of hypotheses, it is in conformity with intuitive expectations: when it comes to the voting decision, drivers are probably more able to personally identify with the situation, thus it is more difficult for them to decide, even though the decision outcome is eventually generally not different from non-drivers. However, it is interesting that a similar intuitive expectation about difference in the difficulty rating was not confirmed for the 1-st person "sitting" scenario.

| Driver flag / Kill whom | Me | Stranger | Total |
|---|---|---|---|
| Non-Driver | 14 | 32 | **46** |
| Active Driver | 33 | 97 | **130** |
| **Total** | **47** | **129** | **176** |

*Table 3: Contingency table of the respondents' active driver flag with respect to the dilemma outcome in the 1st person "sitting" scenario.*

| Driver flag / Kill whom | Driver | Pedestrian | Total |
|---|---|---|---|
| Non-Driver | 10 | 41 | **51** |
| Active Driver | 28 | 108 | **136** |
| **Total** | **38** | **149** | **187** |

*Table 4: Contingency table of the respondents' active driver flag with respect to the dilemma outcome in the 3rd person "voting" scenario.*

To conclude the results related to gender and driver flag, we can sum up that when deciding in the "sitting" scenario, there may be a gender-based bias (it is more difficult for women to decide). When deciding in the "voting" scenario, there may be a driver-based bias (it is more difficult for drivers to decide).

### 4.3 Education

According to our findings, education does not play any crucial role in the decision outcome of the AC dilemmas nor in the perceived level of difficulty of these decisions. This directly contradicts our hypothesis (H5).

### 4.4 Temperament and Character Inventory test and Big Five

The hypotheses (H1) and (H2) are at least partially confirmed by our results as we can show that there are some significant relations between psychological traits and dilemmas outcomes and perceived level of difficulty. In addition, our analysis shows more detailed picture of these associations than expressed in the initial hypotheses. For complete mutual correlation of personality traits see Figure 2.

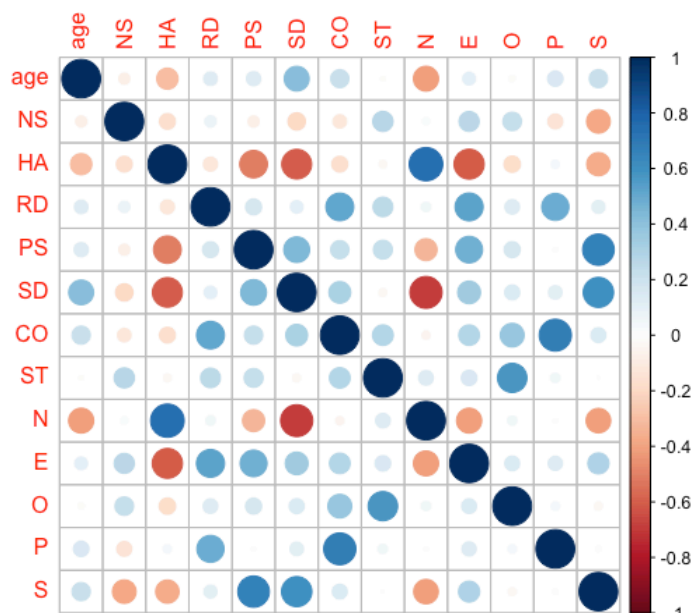

*Figure 2: Mutual correlation of TCI and Big Five variables.*

There is a statistically significant association between Cooperativeness (CO), Persistence (PS) and the decision whom to kill in the 1-st person "sitting" scenario (either "Me", or the "Stranger"). Other TCI test values did not show statistical significance. Logistic regression with CO+PS shows:

- If CO increases by 1.0, the odds of letting the driver kill the stranger/pedestrian decrease by 0.67 (with $p = 0.041$), i.e. the odds of letting the driver kill herself increase appropriately.
- If PS increases by 1.0, the odds of letting the driver kill the stranger/pedestrian increase by 1.44 (with $p = 0.026$), i.e. the odds of letting the driver kill herself decrease appropriately.

These results are in conformity with prior expectations: higher CO is associated with how much people identify with and accept others; higher PS generally means that people are less likely to give up in spite of difficulties, frustration or fatigue, hence probably not give up even in case of an autonomous car accident. The self-transcendence association with the dilemmas outcomes was not confirmed.

There is no statistically significant association between TCI variables or Big Five variables and the decision whom to kill in the 3-rd person "voting" scenario (either "Driver", or the "Pedestrian"). This is an interesting and relevant fact to note: the "sitting" scenario outcome is associated with psychological traits, whereas the "voting" scenario outcome is not. Therefore, there is probably a lower psychological bias in the "voting" scenario.

There is a statistically significant association between Reward Dependence (RD) and the difficulty of the decision in the 1-st person "sitting" scenario (on the scale 1-5). Other TCI test values did not show statistical significance. Ordinal regression with all TCI variables shows that increasing RD by 1.0 (while controlling for all other TCI variables) increases odds of giving a more difficult rating by 1.397 (with $p = 0.022$). This result suggests that even though people with higher RD would decide in the 1-st person "sitting" scenario similarly like other people, this decision is more difficult for them.

There is a statistically significant association between Conscientiousness (S, Svedomitost in Czech) and the decision whom to kill in the 1-st person "sitting" scenario (either "Me", or the "Stranger"). Other Big Five test values did not show statistical significance. Logistic regression with S alone shows that if S increases by 1.0, the odds of letting the driver kill the stranger/pedestrian increase by 1.48 (with $p = 0.035$), i.e. the odds of letting the driver kill herself decrease appropriately. This result is not contradicting prior intuitive expectations because higher S is also associated with being organized and highly planned (even stubborn), as opposed to flexibility, spontaneity and sloppiness associated with low S. However, in our analysis the significance of this result can be weakened by the fact of multiple hypothesis testing and potential data dredging.

There is no significant association between Big Five variables and the 1-st person "sitting" nor the 3-rd person "voting" scenario difficulty rating.

### 4.5 Trolley problem

Contrary to one of the initial hypotheses (H7), there is no significant association between the trolley problem outcome and the autonomous cars dilemma outcome. There is, however, a clearly significant ($p < 0.001$) positive association between the difficulty of answering the first trolley problem question (consequentialists vs. deontologists) and the difficulty of the autonomous cars dilemma in both scenarios. The second trolley problem question difficulty is then not significant.

### 4.6 Predictive analysis

It is not possible to predict the autonomous cars dilemma outcome on the basis of TCI values, Big Five values, age, gender, education and driver flag. Hence, our hypothesis (H8) was not confirmed.

## 4.7 Differences between 1st person and 3rd person scenarios

In spite of the described associations, there is no difference in the mean values of the difficulty rating in the "sitting" vs. "voting" scenarios. This partially rejects one of our initial hypotheses (H6). It is not generally more difficult to decide the "sitting" scenarios. It is only difficult for some groups of respondents, whereas for others it is easier, so the differences cancel out in the mean.

There is not a statistically significant difference ($p = 0.15$) in the probabilities of a pedestrian being killed in the "sitting" vs. "voting" scenarios. In the case of the dataset analysed here, surprisingly a lower percentage of pedestrians (73.3 %) would be killed in the 1-st person "sitting" scenario in comparison with the 3-rd person "voting" scenario (79.7 %). This difference is not significant here.

## 5. Conclusion

Given our empirical data, we have not found a statistically significant difference in the autonomous cars dilemma outcomes between the 1-st person and the 3-rd person scenarios. However, the evidence from our data is still too weak to firmly confirm that no such a difference can exist, and it is still quite likely that if there is a similar survey with a higher number of respondents, the statistical tests will show that there is a significant difference in the true probabilities of pedestrians being killed on the basis of the 1-st person vs. 3-rd person scenario. This hypothesis is also supported by the fact that our pool of respondents was not gender-balanced, comprising less women than men, and by the fact that we have shown that women in comparison to men prefer to get themselves killed more often in the 1-st person scenario. So, if we acquire a larger and more gender-balanced set of responses, we can expect that a higher number of women responding in the 1-st person scenario will actually decrease the percentage of killed pedestrians from current 73 % to an even lower value, whereas the percentage in the 3-rd person scenario will remain about the same (around 80 %) because there is no gender-based difference in the dilemma outcome when using the 3-rd person scenario.

A very practical ethical question arises when an actual accident-handling logic in autonomous cars is to be implemented on the basis of a similarly acquired public consensus. Should we then get to this consensus through the 1-st person scenario leading to more pedestrian-protective cars, or should we rather go through the 3-rd person scenario leading to more passenger-protective cars?

We have shown that the 1-st person scenario outcome and decision difficulty are significantly influenced by several TCI and Big Five psychological meta-factors (Cooperativeness, Persistence, Conscientiousness and Reward Dependence), whereas the 3-rd person scenario outcome is not associated with any TCI and Big Five psychological meta-factors (only its difficulty rating is associated with Cooperativeness). Moreover, gender plays a significant role in the 1-st person scenario outcome as well as difficulty, which is not the case for the 3-rd person scenario where only its difficulty (but not outcome) is associated with the fact whether a respondent is and active driver or non-driver.

This leads us to the conclusion that the 3-rd person scenario is less biased than the 1-st person scenario, and therefore we would recommend any participatory technology assessment discourse to use the 3-rd person scenario, no matter whether it will be eventually shown that it is more passenger-protective or statistically same in terms of passenger protection as the 1-st person scenario.

To put our study in context of research in the ethics and adoption of AC, we consider our results as enriching contribution especially to the research conducted by Costa et al. (2014) that also reveals bias coming from the language and form of communication that is used to collect public opinion on AC. Based on our and Costa's study, we highly recommend to pay attention towards the way any AC related debate is introduced, especially in terms of linguistic and communication aspects.

The impact of psychological traits on a priori adoption of AC has been confirmed by several other studies as well (Kyriakidis et al. 2015, Hohenberger et al. 2017). These findings are interesting especially for marketing strategists and AC industry campaigns since they might be used for influencing consumer behaviour and public opinion, which we should keep in mind and approach very carefully in any future AC policy making.

**References**

Bansal, P. & Kockelman, K. M. (2017) Forecasting Americans' long-term adoption of connected and autonomous vehicle technologies. Transportation Research Part A: Policy and Practice, 95, 49-63.

van den Berg, V. A. C. & Erik T. Verhoef, E. T. (2016) Autonomous cars and dynamic bottleneck congestion: The effects on capacity, value of time and preference heterogeneity. Transportation Research Part B: Methodological. 94, 43-60.

Bonnefon, J.-F., Shariff, A. & Rahwan, I. (2016). The social dilemma of autonomous vehicles. Science 352(6293), 1573–1576.

Cloninger, C. R. (1994). The temperament and character inventory (TCI): A guide to its development and use. St. Louis, Mo: Center for Psychobiology of Personality, Washington University.

Costa, A., Foucart, A., Hayakawa, S., Aparici, M., Apesteguia, J., Heafner, J., & Keysar, B. (2014). Your morals depend on language. PloS One, 9(4), e94842.

Digman, J. M. (1990). Personality structure: emergence of the five-factor model. Annual Review of Psychology, 41(1), 417–440.

Endsley, M. R. (2017) Autonomous driving systems. Journal of Cognitive Engineering and Decision Making. DOI: 10.1177/1555343417695197

Fagnant, D. J., & Kockelman, K. (2015). Preparing a nation for autonomous vehicles: opportunities, barriers and policy recommendations. Transportation Research Part A: Policy and Practice, 77, 167–181.

Fleetwood, J. (2017) Public health, ethics, and autonomous vehicles. American Journal of Public Health, 107(4), 532-537.

Foot, P. (1967). The problem of abortion and the doctrine of double effect. Oxford Review, 5, 5-15.

Geistfeld, M. (2017) A roadmap for autonomous vehicles: state tort liability, automobile insurance, and federal safety regulation. California Law Review, Forthcoming; Research Paper No. 17-09. Available at SSRN: https://ssrn.com/abstract=2931168.

Gerdes, J. C. &Thornton, S. M. (2016) Implementable ethics for autonomous vehicles. In Maurer, M. et al. (Ed.), Autonomous driving: technical, legal and social aspects (pp. 87-102). Springer-Verlag Berlin Heidelberg.

Gogoll, J. & Müller, J.F. (2016) Autonomous cars: in favor of a mandatory ethics setting. Science and Engineering Ethics. DOI:10.1007/s11948-016-9806-x

Goodall, N. J. (2014). Machine Ethics and Automated Vehicles. In Meyer, G. & Beiker, S. (Ed.) Road vehicle automation (pp. 93–102). Springer International Publishing.

Hevelke, A. & Nida-Rümelin, J. (2015). Responsibility for crashes of autonomous vehicles: an ethical analysis. Science and Engineering Ethics, 21(3), 619–630.

Hohenberger, C., Sporlle M., Welpe I.M. (2017) The effects of anxiety on the willingness to use autonomous cars depend on individual levels of self-enhancement. Technological Forecasting & Social Change. 116, 40–52.

ITS (2015) Action Plan for the Deployment of Intelligent Transport Systems (ITS) in the Czech

Republic until 2020 (with the Prospect of 2050). Ministry of Transportation Czech Republic. https://www.dataplan.info/img_upload/7bdb1584e3b8a53d337518d988763f8d/ap-its-zakladni-dokument.pdf. Accessed 25 April 2017.

KPMG (2016) Global automotive executive survey 2016: from a product-centric world to a service-driven digital universe. https://assets.kpmg.com/content/dam/kpmg/pdf/2016/01/gaes-2016.pdf. Accessed 25 April 2017.

Kyriakidis, M., Happee, R. & de Winter, J.C.F. (2015) Public opinion on automated driving: Results of an international questionnaire among 5000 respondents, Transportation Research Part F: Traffic Psychology and Behaviour, 32, 127-140.

Lin, P. (2016). Why ethics matters for autonomous cars. In Maurer, M. et al. (Ed.), Autonomous driving: technical, legal and social aspects (pp. 69–85). Springer-Verlag Berlin Heidelberg.

Milakis, D., van Arem, B. & van Wee, B. (2017): Policy and society related implications of automated driving: a review of literature and directions for future research. Journal of Intelligent Transportation Systems, DOI: 10.1080/15472450.2017.1291351

Navya Technology (2017a) Navya Arma coming into service at Christchurch airport. http://navya.tech/2017/02/navya-arma-coming-into-service-at-christchurch-airport-nz/. Accessed 25 April 2017.

Navya Technology (2017b) UK premiere: Navya Arm shuttle at London's Heathrow airport. http://navya.tech/2017/02/uk-premiere-navya-arma-shuttle-at-londons-heathrow-airport/. Accessed 25 April 2017.

Nevejans, N. (2017) European civil law rules in robotics. Policy Department C: Citizens' Rights and Constitutional Affairs, European Parliament. http://www.europarl.europa.eu/RegData/etudes/STUD/2016/571379/IPOL_STU%282016%29571379_EN.pdf. Accessed 25 April 2017.

Otsuka, M. (2008). Double effect, triple effect and the trolley problem: squaring the circle in looping cases. Utilitas, 20(01), 92–110.

Payre, W., Cestac, J., Delhomme, P. (2014) Intention to use a fully automated car: Attitudes and a priori acceptability. Transportation Research Part F: Traffic Psychology and Behaviour, 27, Part B, 252-263.

PostBus Switzerland Ltd (2017) Shape the mobility of the future. https://www.postauto.ch/smartshuttle. Accessed 25 April 2017.

Rahwan, I., Bonnefon, J-F., Shariff, A., Awad, E., Dsouza, S., Chang, P. & Tang, D. (na) Moral machine. MIT. http://moralmachine.mit.edu. Accessed 25 April 2017.

Ro, Y. & Ha, Y. (2017): A factor analysis of consumer expectations for autonomous cars. Journal of Computer Information Systems, DOI: 10.1080/08874417.2017.1295791

Swann, W. B., Gómez, A., Dovidio, J. F., Hart, S., & Jetten, J. (2010). Dying and killing for one's group: identity fusion moderates responses to intergroup versions of the trolley problem. Psychological Science, 21(8), 1176–1183.

Thomson, J. J. (1985). The trolley problem. The Yale Law Journal, 94(6), 1395-1415.

Volvo Car USA (2016) Survey: New Yorkers and Californians ready for autonomous cars; Texas and Pennsylvania residents sceptical. Volvo Car Corporation.

https://www.media.volvocars.com/us/en-us/media/pressreleases/193745/survey-new-yorkers-and-californians-ready-for-autonomous-cars-texas-and-pennsylvania-residents-skept. Accessed 25 April 2017.

**Appendix**

The full text of the main survey questions (dilemmas scenarios) translated from Czech original to English.

**Trolley problem [version A: switch]**
Imagine yourself in this situation: A train is speeding on a main track. In the distance, you can see five people tied and lying on the same track, unable to move. The train is speeding towards them. You are standing nearby next to a lever. If you pull the lever, the train switches to the side track, where there is another single man tied and lying on the track, unable to move.
**TAQ1: What would you do in that situation?**
(In a real life, the situation might have another solution, but for now imagine that you have just these two options.)
- You rather do nothing and the train is going to kill five people on the main track.
- You pull the lever and the train is going to kill one man in the side track.

**TAQ2: To make that decision about the described situation was for you:**
very easy – somewhat easy – neither easy nor difficult – somewhat difficult – very difficult

**Trolley problem [version B: big man]**
Imagine yourself in this situation: A train is speeding on a main track right towards five people tied and lying on the same track, unable to move. You are watching the situation from the bridge above the track. On the bridge, there is a big man standing next to you, who could surely stop the train with his body.
**TAQ1: What would you do in that situation?**
(In a real life, the situation might have another solution, but for now imagine that you have just these two options.)
- You rather do nothing and the train is going to kill five people on the main track.
- You push the big man from the bridge to the track, where the train will kill him and stop.

**TAQ2: To make that decision about the described situation was for you:**
very easy – somewhat easy – neither easy nor difficult – somewhat difficult – very difficult

**Autonomous cars problem [version A: 1-st person perspective aka "sitting" scenario]**
Imagine yourself in this situation: You are sitting in an autonomous car, which is going on a narrow mountain road, approaching a narrow tunnel. Just before your car enters the tunnel, a man suddenly gets on the road in front of your car, trying to cross it. He/she suddenly slips, falls down and blocks the path.
**AAQ1: What should your autonomous car do in that situation?**
- Run over that human and kill him/her.
- Turn the steering out of the road, hit the wall and kill you.

**AAQ2: To make that decision about the described situation was for you:**
very easy – somewhat easy – neither easy nor difficult – somewhat difficult – very difficult

**Autonomous cars problem [version B: 3-rd person perspective aka „voting" scenario]**

Imagine this situation: An autonomous car is going on a very narrow mountain road with one passenger on board, approaching a narrow tunnel. Just before your car enters the tunnel, a man suddenly gets on the road in front of your car, trying to cross it. He/she suddenly slips, falls down in the middle of the road and blocks the path.

An autonomous car is going on the narrow mountain road approaching the tunnel. At the beginning of the tunnel, there is a human, who tries to cross the road. Just before the car enters the tunnel, a man suddenly gets on the road in front of the car, trying to cross it. He/she suddenly slips, falls down in the middle of the road and blocks the path.

**ABQ1: Imagine that you are voting in a referendum on a new legislation determining how autonomous cars will respond in such situation. What should the autonomous cars do according to your opinion?**
- In such situation, autonomous cars should protect passengers. That means to run over that human and kill him/her.
- In such situation, autonomous cars should protect pedestrians. That means to turn the steering out of the road, hit the wall and kill the passenger.

**ABQ2: To make that decision about the described situation was for you:**
very easy – somewhat easy – neither easy nor difficult – somewhat difficult – very difficult